\documentclass[twocolumn,showpacs,floatfix,superscriptaddress,amsmath,amssymb,prl]{revtex4}
\usepackage{amsfonts}
\usepackage{amsmath}
\usepackage{amssymb}
\usepackage{graphicx}
\usepackage{hyperref}
\usepackage{ulem}

\newcommand{\be}{\begin{equation}}
\newcommand{\ee}{\end{equation}}

\begin{document}

\title{Deriving local order parameters from tensor network representations}

\author{Huan-Qiang Zhou}
\affiliation{Centre for Modern Physics and Department of Physics,
Chongqing University, Chongqing 400044, The People's Republic of
China}

\begin{abstract}
A large class of  quantum phase transitions for quantum lattice
systems are characterized by local order parameters. It is shown
that local order parameters may be systematically constructed from
tensor network representations of quantum many-body ground state
wave functions by investigating the reduced density matrices for
local areas on an infinite-size lattice. Depending on whether or not
the system symmetries are spontaneously broken, and whether or not
the ground state fidelity per lattice site is continuous, there are
four categories of quantum phase transitions for systems with local
order parameters.  Quantum phase transitions characterized by
nonlocal order parameters are discussed, aiming at better
understanding quantum systems with topological order.
\end{abstract}

\date{\today}

\maketitle

Quantum phase transitions (QPTs)~\cite{sachdev} are driven by
quantum fluctuations due to the uncertainty principle, and occur at
zero temperature when some control parameter varies. In the last
decades, much attention has been paid to QPTs in strongly correlated
electronic systems, since some of the most exciting discoveries in
condensed matter physics, such as various magnetic orderings, the
integer and fractional quantum Hall effects, and the high-$T_c$
superconductors, are often attributed to quantum critical phenomena.
In the conventional Landau-Ginzburg-Wilson paradigm, the key notions
are symmetry-breaking orders and local order parameters, with
symmetry-broken phases characterized by the non-zero values of local
order parameters. A new paradigm emerges for systems with
topological order~\cite{wen}, where no local order parameter exists
to characterize exotic orders.

In spite of their decisive role in characterizing QPTs, no scheme,
which is applicable (at least in principle) to quantum systems
undergoing QPTs,  is available to systematically construct order
parameters (either local or nonlocal). The only exception, to the
best of our knowledge, is the work by Furukawa, Misguich and
Oshikawa~\cite{oshikawa}, who attempted to systematically derive
order parameters from comparing the reduced density matrices of the
\textit{degenerate} ground states for various subareas of the system
defined on a \textit{finite-size} lattice. In their approach, the
low-lying energy spectrum and eigenvectors of a finite-size system
with a given discrete symmetry are known (analytically or
numerically), where a finite number of  ground states are assumed to
be nearly degenerate, with their quantum numbers indicating what
symmetries are broken in the thermodynamic limit. Therefore, the
approach, as it stands, only works for systems with a finite number
of (nearly) degenerate ground states.

The difficulties of determining order parameters for a QPT in a
given quantum system lie in the facts that (i) usually it is a
formidable task to compute ground state wave functions and (ii) the
ground state phase diagram is often lacking. However, significant
advances have been made in both the classical simulation of quantum
lattice systems and the determination of ground state phase
diagrams: first, a \textit{tensor network} (TN) representation of
quantum many-body wave functions provides an efficient way to
classically simulate quantum many-body
systems~\cite{MPS,TEBD,TI_MPS,PEPS,iTEBD,iPEPS,mera1,mera2}; second,
two novel approaches to study QPTs have been proposed from a quantum
information perspective, namely \textit{entanglement}
~\cite{preskill,osborne,vidal,korepin,levin,entanglement,review} and
\textit{fidelity} ~\cite{zanardi,zhou,fidelity,zov,topo}. In
Ref.~\cite{zov}, a viable scheme to determine the ground state phase
diagram of a quantum lattice system without prior knowledge of order
parameters was proposed. This was achieved by studying the
singularities in the ground state fidelity per lattice
site~\cite{zhou}, combining with a practical way to compute the
fidelity per lattice site for infinite-size lattice systems using
the TN algorithms for translationally invariant systems
\cite{iTEBD,iPEPS}.

In this paper, we develop a practical scheme to systematically
determine local order parameters for quantum many-body systems on
infinite-size lattices from tensor network representations of the
ground state wave functions by investigating the reduced density
matrices for local areas.  It yields that there are four categories
of QPTs characterized by local order parameters, depending on
whether or not the ground state fidelity per lattice site is
continuous, and whether or not the system symmetries are
spontaneously broken. If no local order parameters exist, then the
system is described by a nonlocal order parameter. The latter is
relevant to QPTs in quantum systems with topological order.

{\it Local and nonlocal order parameters.} Consider a
translationally invariant  infinite-size quantum lattice system $S$
in $D$ spatial dimensions described by a Hamiltonian $H(\lambda)$
with a global symmetry group $G$, where $\lambda$ is a control
parameter~\cite{multiple}. Suppose the system undergos a QPT at a
transition point $\lambda = \lambda_c$. Then any two different
ground states $|\psi(\lambda_1)\rangle$ and
$|\psi(\lambda_2)\rangle$ corresponding to two different values
$\lambda_1$ and $\lambda_2$ of the control parameter $\lambda$ are
orthogonal to each other, i.e., $ \langle
\psi(\lambda_2)|\psi(\lambda_1)\rangle =0$. Therefore, the ground
states  $|\psi(\lambda)\rangle$  for different values of the control
parameter $\lambda$ are reliably distinguishable via quantum
measurements~\cite{nielsen}.  This leads to the existence of
\textit{local} physical observables, which enable to distinguish the
states by means of local measurements, if sufficient copies of the
system are \textit{simultaneously} prepared~\cite{walgate1} for
\textit{different} values of the control parameter. However, a more
pertinent question is to ask if there is any local physical
observable
 $O$ (Hermitian and traceless) defined in a local area
$\Omega$ on the lattice that may tell us in which phase a given
state $|\psi(\lambda)\rangle$ is: $\langle O \rangle \neq 0$ in one
phase and $\langle O \rangle = 0$ in the other phase, if copies of
the system for one \textit{single} value of the control parameter
are prepared~\cite{orderparameter}. If the answer to the question is
affirmative, then we may \textit{define} the local physical
observable as a \textit{local order parameter}~\cite{defination}.
Depending on whether or not the fidelity per lattice site is
continuous~\cite{firstorder}, and whether or not the symmetries are
spontaneously broken, we have four different categories of QPTs with
local order parameters: (1) discontinuous, symmetries not broken;
(2) discontinuous, symmetries broken; (3) continuous, symmetries not
broken; and (4) continuous, symmetries broken. If no local order
parameter exists, then one may further seek a nonlocal physical
observable $O_n$ defined on a nonlocal area on the infinite-size
lattice to judge in which phase a given state
$|\psi(\lambda)\rangle$ is, with only copies of the system prepared
for one \textit{single} value of the control parameter. Such a
nonlocal physical observable $O_n$ is a \textit{nonlocal order
parameter}. This accommodates systems with exotic topological
order~\cite{notall}, which may even coexist with local
symmetry-breaking orders.

{\it Reduced density matrices for local areas on an infinite
lattice.} Now we are in a position to clarify what consequences one
may draw from the existence of a local order parameter $\langle O
\rangle$ on a local area $\Omega$. Here we emphasize that, due to
translational invariance, what really matters is only the size and
shape of the area  $\Omega$, rather than its relative position in
the entire lattice. For our purpose, we partition the whole system
into two parts-the local area $\Omega$ and its complement to the
entire lattice $\bar{\Omega}$. As such, the local area  $\Omega$ is
described by the reduced density matrix $\rho_\Omega(\lambda)$
corresponding to a given ground state $|\psi(\lambda)\rangle$.
Generically, the reduced density matrix $\rho_\Omega(\lambda)$
exhibits nontrivial form due to the distinguishability by local
measurements~\cite{walgate1}, except for the constraints imposed by
the global symmetries (in the symmetric phase). An important fact is
that the reduced density matrix $\rho_\Omega(\lambda)$ possesses
different \textit{structures} in two phases as far as the nonzero
entries are concerned: $\rho_\Omega(\lambda) =\rho_{\Omega
0}(\lambda)$ in one phase and $\rho_\Omega (\lambda)=\rho_{\Omega
0}(\lambda)+ \langle O \rangle O / (\rm {Tr} O^2)$ in the other
phase, as follows from the presence of the local order parameter
$\langle O \rangle$. Two cases should be distinguished: (i) if $O$
is not invariant under the symmetry transformation, then the
symmetry is spontaneously broken; (ii) if $O$ is invariant under the
symmetry transformation, then no symmetry is broken.

We stress that local order parameters are not unique. Indeed,
suppose a local order parameter $\langle O \rangle$ exists for a
local area $\Omega$, then there exists another local order parameter
$\langle \tilde{O}\rangle$ for a larger area $\tilde{\Omega} \supset
\Omega$, since all information encoded in the reduced density matrix
$\rho_\Omega(\lambda)$ should be encoded in the reduced density
matrix $\rho_{\tilde{\Omega}}(\lambda)$, due to the fact that
$\rho_\Omega(\lambda)$ is obtained from
$\rho_{\tilde{\Omega}}(\lambda)$ by tracing out extra degrees of
freedom in the complement set $\tilde{\Omega}/\Omega$. This implies
that there is an \textit{optimal} local order parameter that
corresponds to the smallest local area for the system
considered~\cite{coexistence}.

On the other hand, if the reduced density matrices for all possible
local areas share the \textit{same} nonzero entries structure in two
phases,  then a \textit{nonlocal} order parameter is necessary to
characterize the QPT. That is, an area with nontrivial topology
should be chosen to see if the corresponding reduced density
matrices exhibit different structures in different
phases~\cite{thumbrule}.

{\it Computation of the reduced density matrices from tensor network
representations.} Now we show that it is feasible to compute the
reduced density matrices for various local areas on infinite-size
lattices. In this regard, we rely heavily on the fact that the TN
representation of quantum many-body wave functions provides a
powerful means  to efficiently simulate infinite-size quantum
lattice systems in one~\cite{iTEBD} and two and higher~\cite{iPEPS}
spatial dimensions. In one spatial dimension, the NT representation
is the matrix product states (MPS)~\cite{MPS}, and, in two and
higher spatial dimensions, the NT representation is the projected
entangled-pair states (PEPS)~\cite{PEPS}. In both cases, the TN
algorithms for infinite-size lattices offer an efficient way to
compute the reduced density matrices for various local areas, if
quantum lattice systems are in gapful phases.

If quantum lattice systems are in gapless phases, then a more
sophisticated representation, i.e.,  the multi-scale entanglement
renormalization ansatz (MERA)~\cite{mera1,mera2} is needed for
quantum many-body states. A MERA also provides an efficient
representation of quantum ground states on an infinite-size lattice.
Thanks to the fact that the width of the causal cone is bounded,  a
MERA offers an efficient way to compute the reduced density matrices
for various local areas. In Ref.~\cite{mera1}, it is described
explicitly how to compute the one-site and two-site reduced density
matrices from a MERA.

{\it A practical scheme to derive local order parameters from tensor
network representations.}  For a quantum lattice system with a
symmetry group $G$, one may systematically derive local order
parameters from the TN representations of quantum many body ground
state wave functions. It consists of two steps: (1) determine the
ground state phase diagram by computing the ground state fidelity
per lattice site in terms of the infinite TN algorithms~\cite{zov};
(2) derive local order parameters from a representative ground state
wave functions by investigating the reduced density matrices for
local areas on an infinite-size lattice.

To be self-contained, let us briefly recall the definition of the
ground state fidelity per lattice site $d(\lambda_1,\lambda_2)$. For
any two ground states $|\psi(\lambda_1)\rangle$ and
$|\psi(\lambda_2)\rangle$, the fidelity $F(\lambda_1,\lambda_2)$
asymptotically scales as $F(\lambda_1,\lambda_2) \sim
{d(\lambda_1,\lambda_2)}^N$, with $N$ the total number of sites in
the lattice. Here $d(\lambda_1,\lambda_2)$ is the fidelity per
lattice site~\cite{zhou,zov}, which is well defined in the
thermodynamic limit:
\begin{equation}
d(\lambda_1,\lambda_2) = \lim_{N \rightarrow \infty}
F^{\frac{1}{N}}(\lambda_1,\lambda_2). \label{d}
\end{equation}
It satisfies the properties inherited from the fidelity
$F(\lambda_1,\lambda_2)$: (i) normalization
$d(\lambda_1,\lambda_1)=1$; (ii) symmetry
$d(\lambda_1,\lambda_2)=d(\lambda_2,\lambda_1)$; and (iii) range $0
\le d(\lambda_1,\lambda_2)\le 1$. In fact, the ground state fidelity
$F(\lambda_1,\lambda_2)$ may be mapped onto the partition function
of a $D$-dimensional classical statistical vertex lattice model with
the same geometry~\cite{zov}. Thus, the fidelity per lattice site
$d(\lambda_1,\lambda_2)$ is nothing but the partition function per
site in the classical statistical vertex lattice
model~\cite{baxter}. This justifies why QPTs may be detected as
singularities in $\ln d(\lambda_1,\lambda_2)$ as a function of
$\lambda_1$ and $\lambda_2$.

Once the ground state phase diagram is determined, one may choose a
representative ground state from each phase and investigate the
reduced density matrix for a local area $\Omega$ on an infinite-size
lattice. If the nonzero entries structures of the reduced density
matrices for a given local area $\Omega$ are different for different
phases, then one may read off a local order parameter: (i) if the
reduced density matrices are invariant under the symmetry group $G$,
then no symmetry is spontaneously broken (and the ground state is
non-degenerate); (ii) if one of the reduced density matrices is not
invariant under the symmetry group $G$, then the symmetry is
spontaneously broken (and degenerate ground states arise). If the
nonzero entries structure of the reduced density matrices for all
possible local areas $\Omega$ is the same for different phases, then
there is no local order parameter~\cite{topology}, and vice versa.
For systems with topological order, this is consistent with the fact
that different states in the ground state subspace share the same
bulk tensors in the TN representations, with the only difference at
the top tensor~\cite{topologicalorder}.

{\it Examples.} Let us give a few examples to illustrate our general
scheme. The first example is the two-dimensional quantum Ising model
described by the Hamiltonian:
\begin{equation}
H= -\sum_{(\vec{r},\vec{r}')}  \sigma^{[\vec{r}]}_z \sigma^{[\vec{r}']}_z  -\lambda \sum_{\vec{r}} \sigma^{[\vec{r}]}_x - \epsilon \sum_{\vec{r}} \sigma^{[\vec{r}]}_z
. \label{HXY}
\end{equation}
Here $\sigma^{[\vec{r}]}_x$ and $\sigma^{[\vec{r}]}_z$ are the Pauli
matrices at the lattice site $\vec{r}$, and the control parameters
$\lambda$ and $\epsilon$ correspond to transverse and parallel
magnetic fields. Note that if $\epsilon=0$, then the model enjoys
the $Z_2$ symmetry; otherwise no symmetry exists. As shown in
Ref.~\cite{zov}, for $\epsilon=0$, the fidelity per site
$d(\lambda_1,\lambda_2)$ is continuous, but it exhibits a \textit
{pinch point} singularity, indicating a continuous phase transition
at $\lambda_c \approx 3.044$~\cite{blote}.  On the other hand, the
one-site reduced density matrix $\rho_1$ displays different
nonzero-entries structures in two phases: $\rho_1^{[\vec{r}]} = 1/2+
1/2 \langle \sigma^{[\vec{r}]}_x \rangle  \sigma^x + 1/2 \langle
\sigma^{[\vec{r}]}_z \rangle  \sigma^z$, with $\langle
\sigma^{[\vec{r}]}_z \rangle$ being zero for $\lambda >\lambda_c$
and nonzero for $\lambda <\lambda_c$, as evaluated using the
infinite TN algorithm in Ref.~\cite{iPEPS}.   $\rho_1$ is nontrivial
under the global $Z_2$ symmetry transformation, so the $Z_2$
symmetry is spontaneously broken.  This implies the existence of a
local order parameter: $\langle O \rangle = \langle
\sigma^{[\vec{r}]}_z \rangle$. For nonzero $\epsilon$, when
$\lambda<\lambda_c$, the fidelity per site
$d(\epsilon_1,\epsilon_2)$ is discontinuous~\cite{zov}, implying
that a first order QPT occurs when $\epsilon$ changes sign. The
one-site reduced density matrix $\rho_1$ displays different
structures in different phases: $\langle \sigma^{[\vec{r}]}_z
\rangle$ changes sign, since the symmetry group is
trivial~\cite{orderparameter}.

The second example is a spin 1/2 model with three-body interactions:
\begin{equation}
H= \sum_i 2(g^2 -1) \sigma^z_i \sigma^z_{i+1} - (1+g)^2 \sigma^x_i +
(g-1)^2 \sigma^z_i \sigma^x_{i+1} \sigma^z_{i+2}.
\end{equation}
It is $Z_2$-symmetric under the global spin reversal in the $z$
direction.  As emphasized in Ref.~\cite{zhou}, the parameter space
should be compactified by identifying $g=+\infty$ with $g=-\infty$,
due to the fact that $H(+ \infty) =H(-\infty)$. Since the ground
state is an MPS~\cite{wolf}, one may extract the fidelity per site
$d$ as $d(g,g') = \sqrt {1+|g g'|}/ \sqrt {(1+|g|)(1+|g'|)}$ if $g$
and $g'$ are in different phases, and $d(g,g') = (1+\sqrt {|g g'|})/
\sqrt {(1+|g|)(1+|g'|)}$ if $g$ and $g'$ are in the same phase. Thus
there are two transition points: $g=0$ and $\infty$. All states for
positive $g$ flow to the product state ($g=1$) with all spins
aligning in the $x$ direction, and all states for negative $g$ flow
to the cluster state~\cite{raussendorf} ($g=-1$). Since the ground
state is unique, so no symmetry is spontaneously broken. Therefore,
the reduced density matrices should be invariant under the $Z_2$
symmetry group. Actually, the one-site reduced density matrix
$\rho_1$ exhibits different nonzero entries structures in different
phases: $\rho_1 (i) = 1/2 + 1/2 \langle \sigma^x_i \rangle
\sigma^x$, with $\langle \sigma^x_i \rangle = 4 g/ (1+g)^2$ for $g
>0$ and $\langle \sigma^x_i \rangle = 0$ for $g <0$, thus leading to
the local order parameter $\langle \sigma^x_i \rangle$.

The third example is the spin 1 $XXZ$ model with uniaxial single-ion
anisotropy described by the Hamiltonian:
\begin{equation}
H= \sum_i \left(S^x_i S^x_{i+1}+ S^y_i S^y_{i+1}+J_z S^z_i S^z_{i+1}\right) +D \sum_i {S_i^z}^2,
\end{equation}
where $S^\alpha_i\;(\alpha =x,y,z)$ are the spin 1 operators at the
lattice site $i$, and $J_z$ and $D$ are the Ising-like and
single-ion anisotropies, respectively. The model exhibits rich
phases~\cite{schulz,dennijs,boschi,chen}. For $J_z >0$, three gapful
phases, i.e., the large-$D$, the Haldane, and the N\'eel phases
occur, which have the symmetric, fully broken and partially broken
$Z_2 \times Z_2$ symmetry, respectively. It has been argued that
there is a tricritical point $(J_t,D_t) \approx
(3.20,2.90)$~\cite{boschi,chen}. If $J_z < J_t $, then there are two
critical values $D_{c_1}$ and $D_{c_2}$ when $D$ varies from
$-\infty$ to $\infty$, characterizing the Ising-like transition from
the N\'eel phase to the Haldane phase and the Gaussian transition
from the Haldane phase to the large-$D$ phase, respectively.  Beyond
the tricritical point where  the Haldane phase disappears, the
large-$D$-N\'eel transition was believed to be first order, although
no final proof is available~\cite{chen}. This has been confirmed
numerically by evaluating the fidelity per site based on the
infinite TN algorithm~\cite{li}.  The one-site reduced density
matrix $\rho_1$ exhibits different nonzero-entries structures in
different phases: it is invariant under the spin reversal $Z_2$
symmetry in the large-$D$ phase, but not in the N\'eel phase. This
indicates that the $Z_2$ symmetry is spontaneously broken, with a
local order parameter $\langle O \rangle =(-1)^i \langle S^z_i
\rangle$. No local order parameter exists in the Haldane
phase~\cite{string}.

Support from the Natural Science Foundation of China  is
acknowledged.


\begin{thebibliography}{99}
\bibitem{multiple} It is straightforward to extend the discussion to a system
depending on more than one control parameters.
\bibitem{walgate1} For a quantum system in finite dimensions, if one copy of the
quantum system is prepared in one of  two orthogonal ground states
$|\psi(\lambda_1)\rangle$ and $|\psi(\lambda_2)\rangle$, then
$|\psi(\lambda_1)\rangle$ may always be distinguished from
$|\psi(\lambda_2)\rangle$ by means of local operations and classical
communication (LOCC)~\cite{walgate}. A more refined statement is
that $n$ possible orthogonal states $|\psi(\lambda_i)\rangle
(i=1,\cdots,n)$ can be distinguished reliably with (at most) $n-1$
copies~\cite{walgate}.
\bibitem{orderparameter} Actually, this is valid for local
order parameters in systems with a nontrivial symmetry group $G$.
However, if the symmetry group $G$ is explicitly broken, then a
first-order QPT occurs, due to the presence of the term linear in
$O$, we require that $\langle O \rangle
>0$ in one phase and $\langle O \rangle < 0$ in the other phase, if
only copies of the system are prepared for one \textit{single} value
of the control parameter.
\bibitem{defination} This defination is consistent with the conventional one for symmetry-breaking
orders.
\bibitem{coexistence} This also makes it possible to accomodate
coexisting orders in one phase with more than one order parameters.
\bibitem{thumbrule} There is no simple way to judge if a system
posseses a local or nonlocal order parameter. However, a rule of
thumb for a local order parameter to exist in the system is that if
a product state exists in a given phase (as, e.g., a stable fixed
point in a renormalization group flow), then there must be a local
order parameter. On the other hand, topological ordered systems
possess a degenerate ground state subspace, in which different
states lead to the same reduced density matrix for any local
area~\cite{topology}.
\end{thebibliography}
\end{document}